\pdfoutput=1

\documentclass[twocolumn,floatfix]{revtex4}
\usepackage{graphicx}
\usepackage{amssymb,amsmath}
\begin{document}

\title{
  Magnetoconductance of the Corbino disk in graphene
}

\author{Adam Rycerz}
\affiliation{Marian Smoluchowski Institute of Physics, 
Jagiellonian University, Reymonta 4, PL--30059 Krak$\acute{o}$w, Poland}
\affiliation{Institut f\"{u}r Theoretische Physik, 
Universit\"{a}t Regensburg, D--93040 Regensburg, Germany}

\begin{abstract}
Electron transport through the Corbino disk in graphene is studied in the presence of uniform magnetic fields. At the Dirac point, we observe conductance oscillations with the flux piercing the disk area $\Phi_d$, characterized by the period $\Phi_0=2\,(h/e)\ln(R_\mathrm{o}/R_\mathrm{i})$, where $R_\mathrm{o}$ ($R_\mathrm{i}$) is the outer (inner) disk radius. The oscillations magnitude increase with the radii ratio and exceed $10\%$ of the average conductance for $R_\mathrm{o}/R_\mathrm{i}\geqslant 5$ in the case of the normal Corbino setup, or for $R_\mathrm{o}/R_\mathrm{i}\geqslant 2.2$ in the case of the Andreev-Corbino setup. At a finite but weak doping, the oscillations still appear in a limited range of $|\Phi_d|\leqslant\Phi_d^\mathrm{max}$, away from which the conductance is strongly suppressed. At large dopings and weak fields we identify the crossover to a normal ballistic transport regime.
\end{abstract}
\date{\today}
\maketitle

An atomically thin carbon monolayer (graphene) is widely considered as a successor of silicon in future electronic devices \cite{Gei07}. Investigations of the low-energy properties of graphene, governed by the massless Dirac equation, constitute new and thriving sub-area of condensed matter research \cite{Cas09}. Particularly striking feature of clean, undoped graphene samples is that zero density of states is accompanied by a~nonzero, universal value of the conductivity $4e^2/(\pi{h})$ \cite{Kat06a,Two06,Mia07,Dan08}. This is a basic signature of the so-called \emph{pseudodiffusive} regime, in which transport properties of graphene are indistinguishable from those of a classical diffusive conductor \cite{Ben08}. In this regime, the applied magnetic field does not affect the conductivity \cite{Ost06,Lou07} and higher current cumulants \cite{Pra07}. Prada \emph{et al.}\ also show that for high dopings and magnetic fields, the pseudodiffusive behavior is recovered at resonance with the Landau levels (LLs) in the absence of disorder.

\begin{figure}[!hb]
\centerline{\includegraphics[width=0.7\linewidth]{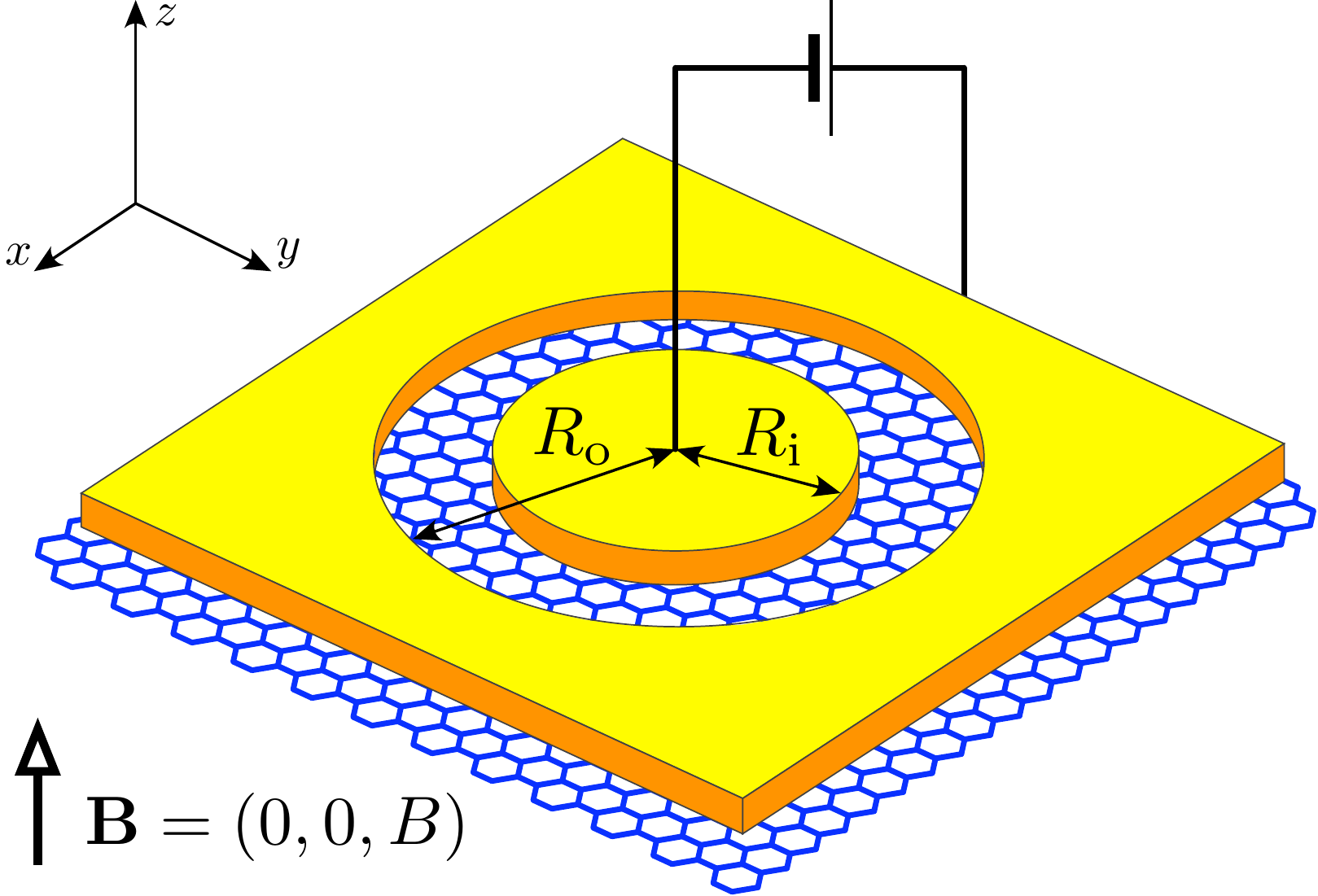}}
\caption{\label{diskset} 
The Corbino magnetometer in graphene. The current is passed through the disk-shaped area with the inner radius $R_\mathrm{i}$ and the outer radius $R_\mathrm{o}$ in a~perpendicular magnetic field $\mathbf{B}=(0,0,B)$. The leads (yellow) are modeled as infinitely-doped graphene regions. The gate electrode (not shown) is placed underneath to tune the doping in the disk area.
}
\end{figure}

Numerous studies of graphene magnetoconductance focus on nanoribbons \cite{LiT09}, Aharonov-Bohm rings \cite{Rus08,Ryc09a}, antidot lattices \cite{She08}, and weak-localization effects in chaotic nanosystems \cite{Tik08,Wur09}. Cheianov and Fal'ko \cite{Che06} showed the conductance of a circular $p$-$n$ interface is insensitive to the weak applied field. The author, Recher, and Wimmer recently identified the crossover from the pseudo-diffusive to the \emph{quantum-tunneling} regime \cite{Ryc09b}, which is characterized by a power-law decay of the conductance $G\propto L^{-\alpha}$ (where $L$ is the length of a sample area and $\alpha$ is a geometry-dependent exponent) and appears for quantum billiard in undoped graphene at \emph{zero} field. In the case of the Corbino disk with the outer radius $R_\mathrm{o}$ and the inner radius $R_\mathrm{i}$ (see Fig.~\ref{diskset}) we have $L=R_\mathrm{o}-R_\mathrm{i}$ and $\alpha=1$, leading to the reciprocal decay of $G$ for $R_\mathrm{o}\gg R_\mathrm{i}$. As the tunneling regime shows up generically for billiards having (at least) one narrow opening \cite{Ryc09b}, the discussion of magnetic field effects---at least on a basic example---is desirable.

In this Rapid Communication, we analyze theoretically magnetoconductance of the Corbino disk in graphene at arbitrary dopings and magnetic fields. The paper is organized as follows: We start from the mode-matching analysis for the disk attached to heavily-doped graphene leads, which employs the total angular momentum conservation in a~similar way as early works employed transverse momentum conservation for the strip geometry \cite{Kat06a,Two06}. Then, we discuss separately the zero- and finite-doping situations, and present the system phase diagram in the field-doping parameter plane. The findings of Refs.\ \cite{Pra07} for the pseudo-diffusive regime are reproduced for $R_\mathrm{o}/R_\mathrm{i}\lesssim 2$. The novel feature is a~periodic (approximately sinusoidal) magnetoconductance oscillation, visible in undoped or weakly doped disks with larger radii ratios, and recovered at LLs for high dopings. Finally, we extend the analysis to the normal-graphene-superconductor (Andreev-Corbino) setup.

The analysis starts from the Dirac Hamiltonian in a~single valley \cite{valfoo}, which is given by
\begin{equation}
\label{dirham}
  H=v_F\left(\mbox{\boldmath$p$}+e\mbox{\boldmath$A$}\right)\cdot\mbox{\boldmath$\sigma$}+U(r),
\end{equation}
where $v_F=10^6\,$m/s is the Fermi velocity, $\mbox{\boldmath$\sigma$}=(\sigma_x,\sigma_y)$, $\mbox{\boldmath$p$}=-i\hbar(\partial_x,\partial_y)$ is the in-plane momentum operator, the electron charge is $-e$, and we choose the symmetric gauge $\mbox{\boldmath$A$}=\frac{B}{2}(-y,x)$. The electrostatic potential energy $U(r)=U_0$ in the disk area ($R_\mathrm{i}<r<R_\mathrm{o}$), otherwise $U(r)=U_\infty$. Since the Hamiltonian (\ref{dirham}) commutes with the total angular momentum operator $J_z=-i\hbar\partial_\varphi+\hbar\sigma_z/2$, the energy eigenfunctions can be chosen as eigenstates of $J_z$
\begin{equation}
\label{jeigen}
  \psi_j(r,\varphi)=e^{i(j-1/2)\varphi}\left(\begin{array}{c}
    \chi_{j\uparrow}(r) \\ \chi_{j\downarrow}(r)e^{i\varphi}
  \end{array}\right),
\end{equation}
where $j$ is an half-odd integer, $s=\uparrow,\downarrow$ denotes the lattice pseudospin, and we have introduced the polar coordinates $(r,\varphi)$. The Dirac equation now reduces to $H_j\chi_j(r)=E\chi_j(r)$,
where $\chi_j(r)=[\chi_{j\uparrow}(r),\chi_{j\downarrow}(r)]^T$, and
\begin{multline}
  H_j=-i\hbar{v_F}\sigma_x\partial_r+U(r)+\\
  \hbar{v_F}\sigma_y\left(\begin{array}{cc}
     \frac{j-1/2}{r}+\frac{eBr}{2\hbar} & 0 \\
     0 & \frac{j+1/2}{r}+\frac{eBr}{2\hbar} \\
  \end{array}\right).
\end{multline}

Subsequently, the scattering problem can be solved separately for each $j$-th angular momentum eigenstate incoming from the origin ($r\!=\!0$). As the angular dependence of the full wavefunction (\ref{jeigen}) does not play a role for the mode matching, the analysis limits effectively to the one-dimensional scattering problem for the spinor $\chi_j(r)$. We model heavily-doped graphene leads by taking the limit of $U(r)=U_\infty\rightarrow\mp\infty$ (hereinafter, the upper sign refers to the conduction band, and the lower sign refers to the valence band), and define the reflection (transmission) amplitudes $r_j$ ($t_j$). For the inner lead ($r<R_\mathrm{i}$), the wavefunction can be written as
\begin{equation}
  \chi_j^{(\mathrm{i})}=\frac{e^{{\pm}ik_{\infty}r}}{\sqrt{r}}
  \left(\begin{array}{c}  1 \\ 1  \end{array}\right)
  +r_j\frac{e^{{\mp}ik_{\infty}r}}{\sqrt{r}}
  \left(\begin{array}{c}  1 \\ -1  \end{array}\right),
\end{equation}
where the first term represents the incoming wave, and the second term represents the reflected wave. We further introduced $k_\infty\equiv|E-U_\infty|/(\hbar{v_F})\rightarrow\infty$. For the outer lead ($r>R_\mathrm{o}$) the wavefunction is
\begin{equation}
  \chi_j^{(\mathrm{o})}=t_j\frac{e^{{\pm}ik_{\infty}r}}{\sqrt{r}}
  \left(\begin{array}{c}  1 \\ 1  \end{array}\right),  
\end{equation}
and represents the transmitted wave. Defining $k_0\equiv|E-U_0|/(\hbar{v_F})$ for the disk area ($R_\mathrm{i}<r<R_\mathrm{o}$), we write the wavefunction in a~similar form as considered by Recher \emph{et al.}\ \cite{Rec09} for the eigenvalue problem, namely
\begin{equation}
  \chi_j^{(\mathrm{d})}=A_j\left(\begin{array}{c}\xi_{j\uparrow}^{(1)} \\ 
     \pm iz_{j,1}\xi_{j\downarrow}^{(1)}\end{array}\right)+
  B_j\left(\begin{array}{c} \xi_{j\uparrow}^{(2)} \\ 
    \pm iz_{j,2}\xi_{j\downarrow}^{(2)}\end{array}\right),
\end{equation}
where $z_{j,1}=[2(j+s_j)]^{-2s_j}$, $z_{j,2}=2(\beta/k_0^2)^{s_j+1/2}$ (with $s_j\equiv\frac{1}{2}\mbox{sgn}\,j$, $\beta=eB/(2\hbar)$), and
\begin{equation}
\label{xisnu}
  \xi_{js}^{(\nu)}=e^{-\beta{r}^2/2}(k_0r)^{|l_s|}\left\{\begin{array}{cc}
    M(\alpha_{js},\gamma_{js},\beta{r}^2), & \nu\!=\!1, \\ 
    U(\alpha_{js},\gamma_{js},\beta{r}^2), & \nu\!=\!2, \\
  \end{array}\right.
\end{equation}
with $l_s=j\mp\frac{1}{2}$ for $s=\uparrow,\downarrow$, $\alpha_{js}=\frac{1}{4}[2(l_{-s}+|l_s|+1)-k_0^2/\beta]$, and $\gamma_{js}=|l_s|+1$. $M(a,b,z)$ and $U(a,b,z)$ are the confluent hypergeometric functions \cite{Abram}. Solving the matching conditions  $\chi_j^{(\mathrm{i})}(R_\mathrm{i})=\chi_j^{(\mathrm{d})}(R_\mathrm{i})$ and $\chi_j^{(\mathrm{o})}(R_\mathrm{o})=\chi_j^{(\mathrm{d})}(R_\mathrm{o})$ we find the transmission probability for $j$-th mode
\begin{equation}
\label{tjdop}
  T_j=|t_j|^2=\frac{16\,(k_0^2/\beta)^{|2j-1|}}{k_0^2R_\mathrm{i}R_\mathrm{o}%
  \,(X_j^2+Y_j^2)}
  \left[\frac{\Gamma(\gamma_{j\uparrow})}{\Gamma(\alpha_{j\uparrow})}\right]^2,
\end{equation}
where $\Gamma(z)$ is the Euler Gamma function, and
\begin{align}
& X_j = w_{j\uparrow\uparrow}^- + z_{j,1}z_{j,2}w_{j\downarrow\downarrow}^-,\ \ \ 
  Y_j = z_{j,2}w_{j\uparrow\downarrow}^+ - z_{j,1}w_{j\downarrow\uparrow}^+, \nonumber\\
& w_{jss'}^\pm= \xi_{js}^{(1)}(R_\mathrm{i})\xi_{js'}^{(2)}(R_\mathrm{o})
  \pm \xi_{js}^{(1)}(R_\mathrm{o})\xi_{js'}^{(2)}(R_\mathrm{i}).
\end{align}
Without loss of generality, we choose $B>0$. For $B<0$ one gets $T_j(B)=T_{-j}(-B)$.

First, we consider the zero doping limit, for which Eq.\ (\ref{tjdop}) simplifies to
\begin{equation}
\label{tjdisk}
  T_j(k_0\!\rightarrow\!0)=
  \frac{1}{\cosh^2\left[\mathcal{L}(j+\Phi_d/\Phi_0)\right]},
\end{equation}
where $\mathcal{L}=\ln(R_\mathrm{o}/R_\mathrm{i})$, $\Phi_d=\pi(R_\mathrm{o}^2-R_\mathrm{i}^2)B$ is the flux piercing the disk area, and $\Phi_0= 2\,(h/e)\mathcal{L}$. {We observe varying the ratio $\Phi_d/\Phi_0$ affects $T_j(k_0\!\rightarrow\!{0})$ similarly as changing boundary conditions affects the corresponding formula for the strip geometry \cite{Kat06a,Two06}.} (Notice that Eq.\ (\ref{tjdisk}) is insensitive to the flux piercing the \emph{inner} lead.) The disk conductance follows by summing over the modes
\begin{equation}
\label{gsdisk}
  G=g_0\sum_{j}T_j(k_0\!\rightarrow\!{0})=
  \sum_{m=0}^\infty G_m\cos\left(\frac{2\pi{m}\Phi_d}{\Phi_0}\right),
\end{equation}
where $g_0=4e^2/h$ is the conductance quantum (the factor $4$ includes spin and valley degeneracy), and the Fourier amplitudes are
\begin{equation}
\label{gscoef}
  G_0=\frac{2g_0}{\mathcal{L}}, \ \ \ \ \ 
  G_{m}=\frac{4\pi^2(-)^mmg_0}{\mathcal{L}^2\sinh(\pi^2m/\mathcal{L})}
  \ \ \ (m\!>\!{0}).
\end{equation}
The conductance given by Eq.\ (\ref{gsdisk}) shows periodic oscillations with the average value $G_0$ equal to the pseudo-diffusive disk conductance \cite{Ryc09b}. {(Thus, the averaging over $\Phi_d$ for the disk corresponds to the fictitious \emph{averaging over boundary conditions} for the strip.)} The approximate formula $G\approx G_0+G_1\cos(2\pi\Phi_d/\Phi_0)$ reproduces the full expression with the $1\%$ accuracy for $R_\mathrm{o}/R_\mathrm{i}\leqslant 10$. The oscillations magnitude $\Delta G\equiv G_\mathrm{max}-G_\mathrm{min}\approx 2|G_1|$ converges rapidly to $0$ with $R_\mathrm{o}/R_\mathrm{i}\rightarrow 1$ (the pseudo-diffusive transport regime), in agreement with earlier works \cite{Ost06,Lou07,Pra07} reporting no field dependence of the conductance. For instance, we obtain $\Delta{G}<4{\cdot}10^{-5}\,G_0$ for $R_\mathrm{o}/R_\mathrm{i}\geqslant 2$. In the tunneling regime, the oscillations magnitude of $\Delta G\gtrsim 0.1G_0$ is reached for moderate radii ratios $R_\mathrm{o}/R_\mathrm{i}\geqslant 5$. In this regime, when $\Phi_d/\Phi_0$ is half-odd integer, the major contribution to the conductance originates from a single mode ($T_{-\Phi_d/\Phi_0}=1$) and we have $G=G_\mathrm{max}$ (with $G_\mathrm{max}\rightarrow g_0$ for $R_\mathrm{o}\gg R_\mathrm{i}$). For other values of $\Phi_d$, the conductance is generally dominated by the two modes, with $j_\pm=-\mbox{int}(\Phi_d/\Phi_0\mp \frac{1}{2})-\frac{1}{2}$, which became equivalent for $\Phi_d/\Phi_0$ integer, when  $G=G_\mathrm{min}$ (and $G_\mathrm{min}R_\mathrm{o}/R_\mathrm{i}\rightarrow 8g_0$ for $R_\mathrm{o}\gg R_\mathrm{i}$) reproducing the zero-field situation \cite{Ryc09b}.

\begin{figure}[!ht]
\centerline{\includegraphics[width=0.8\linewidth]{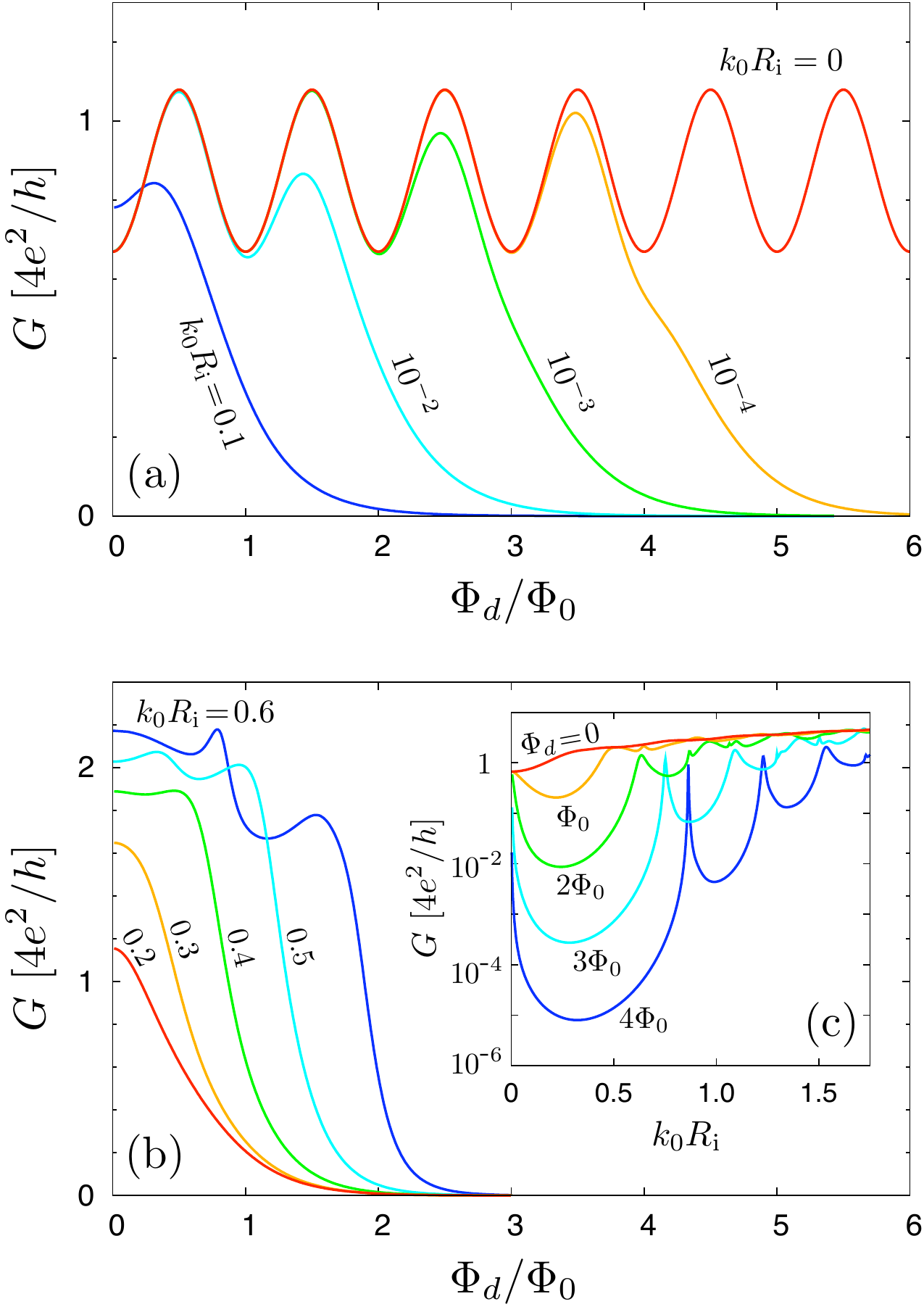}}
\caption{\label{gmag3}
Conductance as a function of the magnetic field and the doping for $R_\mathrm{o}/R_\mathrm{i}=10$. (a) Magnetoconductance at weak doping (specified for each curve by $k_0R_\mathrm{i}=10^{-4}\div 10^{-1}$, with $k_0=|E-U_0|/\hbar{v}_F$). The zero-doping magnetoconductance is also shown (red line). (b) Magnetoconductance at large doping ($k_0R_\mathrm{i}=0.2\div 0.6$). (c) Conductance as a function of doping at fixed magnetic field (specified by the flux piercing the disk area $\Phi_d=0\div 4\Phi_0$, with $\Phi_0=2\,(h/e)\ln{R_\mathrm{o}/R_\mathrm{i}}$.)
}
\end{figure}

We now complement the discussion by analyzing a finite doping case, to find out how stable are the conductance oscillations when the gate voltage is controlled with a finite precision. For $k_0>0$ Eq.\ (\ref{tjdop}) is well defined for arbitrary $j$ provided that $\frac{1}{4}k_0^2/\beta=(k_0l_B)^2/2\neq n=1,2,\dots$ (LLs) with $l_B=\sqrt{\hbar/eB}$ the magnetic length. In such case, the asymptotic form for large fields is $T_j^{(n)}\approx \cosh^{-2}[\mathcal{L}(j-2n+\Phi_d/\Phi_0)]$, leading to conductance oscillations as obtained above, see Eq.\ (\ref{gsdisk}). In fact, $T_j(k_0\rightarrow 0)$ given by Eq.\ (\ref{tjdisk}) are reproduced for $n=0$, showing the conductance oscillations for an undoped disk can be rationalized in terms of resonant transport through the zero-th LL pinned at the Dirac point.

\begin{figure}[!ht]
\centerline{\includegraphics[width=0.9\linewidth]{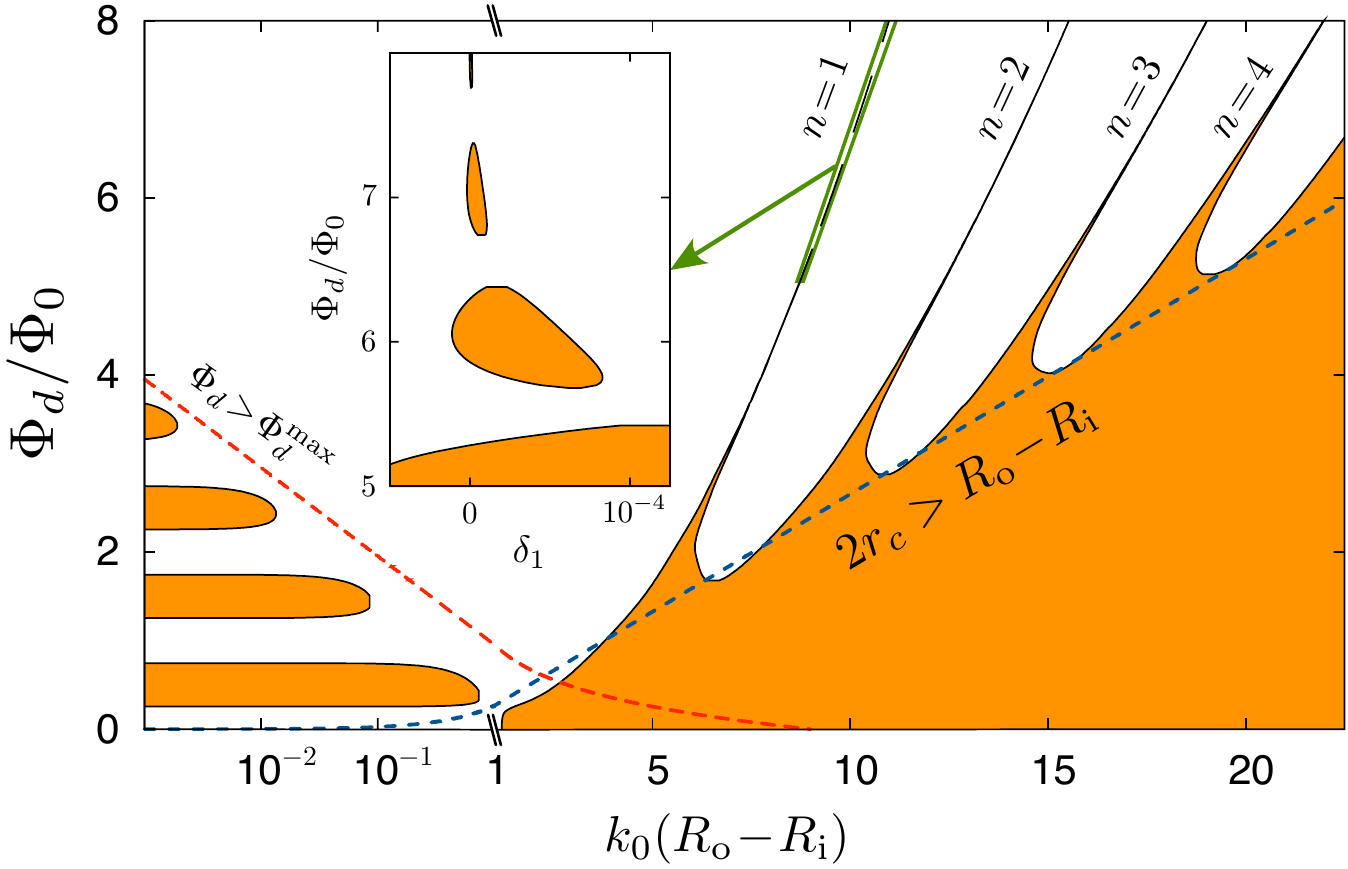}}
\caption{\label{gdiff} 
Phase diagram representing the tunneling, field-suppressed and ballistic transport regimes in the field-doping parameter plane. Solid lines corresponds to $G=G_0$ for the radii ratio $R_\mathrm{o}/R_\mathrm{i}=10$. Dashed lines depict borders of the tunneling ($\Phi_d<\Phi_d^\mathrm{max}$, red line) and ballistic ($2r_c>R_\mathrm{o}-R_\mathrm{i}$, blue line) transport regimes. (Notice the logarithmic scale for $k_0(R_\mathrm{o}\!-\!R_\mathrm{i})<1$.) \emph{Inset} shows the crossover into the tunneling behavior for the first Landau level ($n\!=\!1$) in the magnified horizontal scale ($\delta_n\equiv\frac{1}{2}k_0^2l_B^2-n$). 
}
\end{figure}

The results for $G$, obtained by numerical summation of $T_j$-s given by Eq.\ (\ref{tjdop}) for $R_\mathrm{o}/R_\mathrm{i}=10$, are shown in Fig.\ \ref{gmag3}. We first compare, in Fig.\ \ref{gmag3}(a), the zero-doping magnetoconductance (red line) given by Eq.\ (\ref{gsdisk}) with those obtained for dopings varying from $k_0R_\mathrm{i}=10^{-4}$ to $0.1$, with the steps of one order of magnitude. Weak-doping curves follow the zero-doping one for first few periods, when
\begin{equation}
\label{phdmax}
  \Phi_d\lesssim \Phi_d^\mathrm{max}=\frac{2h}{e}\ln (k_0R_\mathrm{i}).
\end{equation}
For higher fields, $G$ decays as $e^{-(R_\mathrm{o}-R_\mathrm{i})^2/(2l_b^2)}$. The high-$G$ area, limited by Eq.\ (\ref{phdmax}) shrinks rapidly with increasing $k_0$. However, for high dopings ($k_0(R_\mathrm{o}-R_\mathrm{i})\gtrsim\pi$) {the results start to follow the semiclassical picture, similarly as for the two-dimensional electron gas (2DEG) \cite{Kir94}.} The high-$G$ area expands with $k_0$ (see Fig.\ \ref{gmag3}b), as it is now limited by the condition $2r_c>R_\mathrm{o}-R_\mathrm{i}$ (with $r_c=k_0l_B^2$ the cyclotronic radius), characterizing the \emph{ballistic transport} regime. {In particular, for $2r_c>R_\mathrm{o}+R_\mathrm{i}$ we have $G/g_0\approx 2k_0R_\mathrm{i}$, approaching the result for zero field \cite{Ryc09b,stepfoo}.} At high magnetic fields (for which $2r_c<R_\mathrm{o}-R_\mathrm{i}$) we enter the \emph{field-suppressed transport} regime, in which $G\sim e^{-(R_\mathrm{o}-R_\mathrm{i})^2/(2l_b^2)}$ again, except from the isolated peaks (see Fig.\ \ref{gmag3}(c) for the plot in a logarithmic scale), which correspond to the resonances with LLs, and shrink with the field in the absence of disorder. At each resonance, the zero-doping field-dependence of $G$ is approached for the high field. 

The behaviors described above are presented in a~condensed form in the phase diagram shown in Fig.\ \ref{gdiff}. Colored areas represents the regions in the field-doping parameter plane where $G>G_0$ (with the borders $G=G_0$ marked by solid lines). We also show (with dashed lines) the limiting values of the magnetic field, at which the crossovers from the tunneling (left) and from the ballistic (right) to the field-suppressed transport regime occur. For the first LL, we demonstrate in a quantitative manner (see the inset) how, starting from the ballistic regime and enlarging the field (but keeping $(k_0l_B)^2/2\approx 1$), one restores the tunneling behavior, characterized by a chain of isolated islands of $G>G_0$ on the diagram.

So far, we have considered the disk attached to normal-metallic leads. For the Andreev-Corbino setup, with one normal and one superconducting lead, the conductance is expressed in terms of $T_j$-s given by Eq.\ (\ref{tjdop}) as \cite{Akh07}
\begin{equation}
  G^\mathrm{NS}=2g_0\sum_j\frac{T_j^2}{(2-T_j)^2}.
\end{equation}
In particular, $G^\mathrm{NS}$ is still a~periodic function of $\Phi_d$ at the Dirac point, and its Fourier decomposition $G^\mathrm{NS}(\Phi_d)=G_0^\mathrm{NS}+\sum_{m=1}^\infty{G}_m^\mathrm{NS}\cos(2\pi{m}\Phi_d/\Phi_0)$ leads to
\begin{equation}
  G_0^\mathrm{NS}(\mathcal{L})=G_0(\mathcal{L}), \ \ \ \ \ \ 
  G_m^\mathrm{NS}(\mathcal{L})=2G_m(2\mathcal{L}).
\end{equation}
Although we have $G_m^\mathrm{NS}/G_m\rightarrow{1}$ for $R_\mathrm{o}/{R}_\mathrm{i}\rightarrow\infty$ (and any $m$), magnetoconductance oscillations are noticeably amplified for moderate radii ratios. For instance, the magnitudes $\Delta G/G_0>0.1$ are now reached for $R_\mathrm{o}/{R}_\mathrm{i}\geqslant 2.2$. At finite dopings, Eq.\ (\ref{phdmax}) for $\Phi_d^\mathrm{max}$ holds true, and the phase diagram in the field-doping parameter plane (Fig.\ \ref{gdiff}) is almost unaffected. We further notice that for available ballistic graphene samples $2R_\mathrm{o}<1\,\mu$m, and the critical field $B_c$ typically corresponds to $\Phi_d<\Phi_0$. In effect, the zero-field conductance minimum is expected to be significantly deeper than the other minima, for which both electrodes are driven into the normal state. 

In conclusion, we have identified the new transport phenomenon in undoped graphene, which manifests itself by periodic magnetoconductance oscillations for the Corbino geometry. The relative field-induced conductance change reaches experimentally accessible magnitudes $\Delta G/G_0>0.1$ for moderate radii ratios. At weak doping, the oscillations remain observable for a finite range of applied fields \cite{expfoo}. Additionally, we have presented the complete phase diagram in a field-doping parameter plane, illustrating the crossover from the field-suppressed to the ballistic transport regime, as well as the resonances through Landau levels, at which the oscillatory behavior is restored. 

We hope our analysis shall rise some interest in Corbino measurements within the graphene community. {Although the discussion is limited to the system with a perfect circular symmetry and the uniform field, particular features of the results, including (i) the conductance dependce on the \emph{total} flux piercing the sample area and (ii) the formal analogy between dimensionless flux $\Phi_d/\Phi_0$ and the boundary conditions at zero field, suggest magnetoconductance oscillations should appear in more general situation as well.} The work primarily focuses on graphene, but the recent study on effective Dirac fermion model for HgTe/CdTe quantum wells \cite{Sch09} suggests that our findings may also be relevant to such systems.

I thank K.~Richter, P.~Recher, and J.~Wurm for discussions.
The support from the Alexander von Humboldt Stiftung-Foundation and the Polish Ministry of Science (Grant No.\ N--N202--128736) is acknowledged. 

\paragraph*{Note added in proof.} Recently, I became aware of a work on zero-doping Corbino magnetoconductance in graphene employing the conformal mapping technique, Ref.\ \cite{Kat10}.


\begin{thebibliography}{99}
\bibitem{stepfoo}
The quantization steps predicted for 2DEG \cite{Kir94} are missing in the data of Fig.\ \ref{gmag3}(b,c). 
For a detailed discussion of this discrepancy, see Sec.\ III of 
Ref.\ \cite{Ryc09b}.
\end{thebibliography}
\end{document}